\documentstyle[preprint, aps, epsfig]{revtex}
\begin{document}
\textwidth=162mm
\textheight=235mm\def\baselinestretch{1.2}
\global\arraycolsep=2pt
\makeatletter
\def\fmslash{\@ifnextchar[{\fmsl@sh}{\fmsl@sh[0mu]}}
\def\fmsl@sh[#1]#2{%
  \mathchoice
    {\@fmsl@sh\displaystyle{#1}{#2}}%
    {\@fmsl@sh\textstyle{#1}{#2}}%
    {\@fmsl@sh\scriptstyle{#1}{#2}}%
    {\@fmsl@sh\scriptscriptstyle{#1}{#2}}}
\def\@fmsl@sh#1#2#3{\m@th\ooalign{$\hfil#1\mkern#2/\hfil$\crcr$#1#3$}}
\makeatother
\preprint{\vbox{\baselineskip 16pt
\hbox{ASITP-2000-011}
\hbox{SNUTP 00-027}
}}
\title{
Leading Isgur-Wise form factor of $\Lambda_b\to\Lambda_{c1}$ 
transition using QCD sum rules
}
\author{
Ming-Qiu Huang$^{a,b,d}$\footnote{E-mail address:~mqhuang@nudt.edu.cn}, 
Jong-Phil Lee$^c$\footnote{E-mail address:~jplee@phya.snu.ac.kr},
Chun Liu$^d$\footnote{E-mail address:~liuc@itp.ac.cn},
H. S. Song$^c$\footnote{E-mail address:~hssong@physs.snu.ac.kr}}
\address{$^a$CCAST, World Lab., Beijing, 100080, China}
\address{
$^b$Department of Applied Physics, Changsha Institute of Technology,
Hunan 410073, China} 
\address{$^c$School of Physics, Seoul National University, 
Seoul, 151-742, Korea}
\address{
$^d$Institute of Theoretical Physics, Chinese Academy of Sciences, 
P.O. Box 2735,\\ Beijing 100080, China}
\maketitle
\begin{abstract}
The leading Isgur-Wise form factor $\xi(y)$ parametrizing the semileptonic 
transitions $\Lambda_b\to\Lambda^{1/2}_{c1}\ell\bar\nu$ and 
$\Lambda_b\to\Lambda^{3/2}_{c1}\ell\bar\nu$ is calculated by using the QCD 
sum rules in the framework of heavy quark effective theory, where 
$\Lambda^{1/2}_{c1}$ and $\Lambda^{3/2}_{c1}$ is the orbitally excited charmed 
baryon doublet with $J^P=(1^-/2,3^-/2)$.  The interpolating currents with 
transverse covariant derivative are adopted for $\Lambda^{1/2}_{c1}$ and 
$\Lambda^{3/2}_{c1}$ in the analysis.  The slope parameter $\rho^2$ in linear 
approximation of the Isgur-Wise function is obtained to be $\rho^2=2.01$, and 
the interception to be $\xi(1)=0.29$.  The decay branching ratios are estimated.  
\end{abstract}
\pagebreak

\section{Introduction}

The ground state bottom baryon $\Lambda_b$ weak decays \cite{PDG} provide a 
testing ground for the standard model (SM).  They reveal some important 
features of the physics of bottom quark.  The experimental data on them are 
accumulating, and waiting for reliable theoretical calculations.  The main 
difficulties in the SM calculations, however, are due to the poor understanding 
of the nonperturbative aspects of the strong interaction (QCD).  The heavy 
quark effective theory (HQET) based on the heavy quark symmetry provides a 
model-independent method for analyzing heavy hadrons containing a single heavy 
quark \cite{HQET}.  It allows us to expand the physical quantity in powers of 
$1/m_Q$ systematically, where $m_Q$ is the heavy quark mass.  Within this 
framework, the classification of the $\Lambda_b$ exclusive weak decay form 
factors has been simplified greatly.  The decays 
$\Lambda_b\to\Lambda_c l{\bar\nu}$ \cite{IW}, 
$\Lambda_b\to\Sigma_c^{(*)}l{\bar\nu}$ \cite{Mannel},
$\Lambda_b\to\Sigma_c^{(*)} \pi l{\bar\nu}$ \cite{Cho}, 
$\Lambda_b\to p(\Lambda)$ \cite{Qiao} have been studied.

With the discovery of the orbitally excited charmed baryons $\Lambda_c(2593)$ 
and $\Lambda_c(2625)$ \cite{CLEO}, it would be interesting to investigate the 
$\Lambda_b$ semileptonic decays into these baryons.  From the phenomenological 
point of view, these semileptonic transitions are interesting, since in 
principle they may account for a sizeable fraction of the inclusive 
semileptonic rate of $\Lambda_b$ decay.

The properties of excited  baryons have attracted attention in recent years.  
Investigation on them will extend our ability in the application of QCD.  It 
can also help us foresee other excited heavy baryons undiscovered yet.  The 
heavy quark symmetry \cite{HQET} is a useful tool to classify the hadronic 
spectroscopy containing a heavy quark $Q$.  In the infinite mass limit, the 
spin and parity of the heavy quark and that of the light degrees of freedom are 
separately conserved. Coupling the spin of light degrees of freedom $j_\ell$ 
with the spin of heavy quark $s_Q=1/2$ yields a doublet  with total spin 
$J=j_\ell\pm 1/2$ (or a singlet if $j_\ell=0$). This classification can be 
applied to the $\Lambda_Q$-type baryons. For the charmed baryons the ground 
state $\Lambda_c$ contains light degrees of freedom with spin-parity 
$j_\ell^{P}=0^+$, being a singlet. The excited states with $j_\ell^P=1^-$ are 
spin symmetry doublet with $J^P$($1^-/2$,$3^-/2$). The lowest states of such 
excited charmed states, $\Lambda^{1/2}_{c1}$ and $\Lambda^{3/2}_{c1}$, have 
been observed and are identified with $\Lambda_c(2593)$ and $\Lambda_c(2625)$ 
respectively \cite{CLEO}.

The semileptonic $\Lambda_b$ decay rate to the excited charmed baryon is 
determined by corresponding hadronic matrix elements of the weak axial-vector 
and vector currents.  The matrix elements of the vector and axial currents 
($V_\mu={\bar c}\gamma_\mu b$ and $A_\mu={\bar c}\gamma_\mu\gamma_5 b$) between 
the $\Lambda_b$ and $\Lambda^{1/2}_{c1}$ or $\Lambda^{3/2}_{c1}$ can be
parametrized in terms of fourteen form factors:
\begin{mathletters}\label{formfactor}
\begin{eqnarray}
\frac{\langle\Lambda_{c1}^{1/2}(v^\prime,s^\prime)|V_\mu|\Lambda_b(v,s)\rangle}
{\sqrt{4M_{\Lambda_{c1}(1/2)}M_{\Lambda_b}}}&=&
{\bar u}_{\Lambda_{c1}}(v^\prime,s^\prime)
 \Big[F_1\gamma_\mu+F_2 v_\mu+F_3 v^\prime_\mu\Big]
\gamma_5 u_{\Lambda_b}(v,s)~,\\
\frac{\langle\Lambda_{c1}^{1/2}(v^\prime,s^\prime)|A_\mu|\Lambda_b(v,s)\rangle}
{\sqrt{4M_{\Lambda_{c1}(1/2)}M_{\Lambda_b}}}&=&
{\bar u}_{\Lambda_{c1}}(v^\prime,s^\prime)
 \Big[G_1\gamma_\mu+G_2 v_\mu+G_3 v^\prime_\mu\Big]
u_{\Lambda_b}(v,s)~,\\
\frac{\langle\Lambda_{c1}^{3/2}(v^\prime,s^\prime)|V_\mu|\Lambda_b(v,s)\rangle}
{\sqrt{4M_{\Lambda_{c1}(3/2)}M_{\Lambda_b}}}&=&
{\bar u}^\alpha_{\Lambda_{c1}}(v^\prime,s^\prime)\Big[
 v_\alpha(K_1\gamma_\mu+K_2 v_\mu+K_3 v^\prime_\mu)+K_4 g_{\alpha\mu}\Big]
u_{\Lambda_b}(v,s)~,\\
\frac{\langle\Lambda_{c1}^{3/2}(v^\prime,s^\prime)|A_\mu|\Lambda_b(v,s)\rangle}
{\sqrt{4M_{\Lambda_{c1}(3/2)}M_{\Lambda_b}}}&=&
{\bar u}^\alpha_{\Lambda_{c1}}(v^\prime,s^\prime)\Big[
v_\alpha(N_1\gamma_\mu+N_2 v_\mu+N_3 v^\prime_\mu)+N_4 g_{\alpha\mu}\Big]
\gamma_5 u_{\Lambda_b}(v,s)~,
\end{eqnarray}
\end{mathletters}
where $v(v')$ and $s(s')$ are the four-velocity and spin of 
$\Lambda_b(\Lambda_{c1})$, respectively.  And the form factors $F_i$, $G_i$, 
$K_i$ and $N_i$ are functions of $y=v\cdot v'$.  In the limit $m_Q\to\infty$,
all the form factors are related to one independent universal form factor
$\xi(y)$ called Isgur-Wise (IW) function \cite{Choi}.  Extensive investigation 
in \cite{Leibovich} further shows that the leading $1/m_Q$ correction of the 
form factors at zero recoil can be calculated in a model-independent way in 
terms of the masses of charmed baryon states.  A convenient way to evaluate 
hadronic matrix elements is by introducing interpolating fields in HQET 
developed in Ref.~\cite{Falk} to parametrize the matrix elements in 
Eqs. (\ref{formfactor}).  With the aid of this method the matrix element can be 
written as \cite{Leibovich}
\begin{equation}
{\bar c}\Gamma b={\bar h}^{(c)}_{v^\prime}\Gamma h^{(b)}_v
=\xi(y)v_\alpha{\bar\psi}^\alpha_{v^\prime}\Gamma\psi_v~
\label{current}
\end{equation}
at leading order in $1/m_Q$ and $\alpha_s$, where $\Gamma$ is any collection of 
$\gamma$-matrices. The ground state field, $\psi_v$, destroys the $\Lambda_b$ 
baryon with four-velocity $v$; the spinor field $\psi^\alpha_v$ is given by
\begin{equation}
\psi^\alpha_v=\psi^{3/2\alpha}_v
 +\frac{1}{\sqrt{3}}(\gamma^\alpha+v^\alpha)\gamma_5\psi^{1/2}_v~,
\end{equation}
where $\psi^{1/2}_v$ is the ordinary Dirac spinor and $\psi^{3/2\alpha}_v$ is 
the spin 3/2 Rarita-Schwinger spinor, they destroy $\Lambda^{1/2}_{c1}$ and 
$\Lambda^{3/2}_{c1}$ baryons with four-velocity $v$, respectively. To be 
explicit,
\begin{eqnarray}
F_1&=&\frac{1}{\sqrt{3}}(y-1)~\xi(y)~,~~~
G_1=\frac{1}{\sqrt{3}}(y+1)~\xi(y)~,\nonumber\\
F_2&=&G_2=-\frac{2}{\sqrt{3}}~\xi(y)~,~~~~~
K_1=N_1=\xi(y)~,\nonumber\\
&&~~~~~~~~~~~~~~~~~({\rm others})=0~.
\label{formfactors}
\end{eqnarray}
In general, the IW form factor is a decreasing function of the four velocity
transfer $y$.  Since the kinematically allowed region of $y$ for heavy to 
heavy transition is very narrow around unity,
\begin{equation}
1\le y\le 
\frac{M_{\Lambda_b}^2+M_{\Lambda_{c1}}^2}{2M_{\Lambda_b}M_{\Lambda_{c1}}}
\simeq 1.2~,
\end{equation}
it is convenient to approximate the IW function linearly
\begin{equation}
\xi(y)=\xi(1)(1-\rho^2(y-1))~,
\label{rho}
\end{equation}
where $\rho^2$ is the slope parameter which characterizes the shape of the
IW function.

To obtain detailed predictions for the hadrons, however, some nonperturbative 
QCD methods are required.  We adopt QCD sum rules \cite{sumrule0} in this work.
QCD sum rule is a powerful nonperturbative method based on QCD \cite{sumrule0}.
It takes into account the nontrivial QCD vacuum, parametrized by various
vacuum condensates, to describe the nonperturbative nature.  In QCD sum rule, 
hadronic observables are calculable by evaluating two- or three-point 
correlation functions.  The hadronic currents for constructing the correlation 
functions are expressed by the interpolating fields.  The static properties of 
$\Lambda_b$ and $\Lambda_{c1}$ ($\Lambda_{c1}$ denotes the generic 
$j_\ell^{P}=1^-$ charmed state) have been studied with QCD sum rules in the 
HQET in \cite{sumrule} and \cite{PLB476,Zhu}, respectively.  The aim of this 
work is to calculate the leading IW function $\xi(y)$ using the QCD sum rules.

In the next Section, the QCD sum rule calculations for $\xi(y)$ are given.
Numerical results and discussions are in Sec.\ III.  Summary of this work 
is in Sec.\ IV.

\section{The QCD Sum Rule Calculations}

As a starting point, consider the interpolating field of heavy baryons.  The 
heavy baryon current is generally expressed as 
\begin{equation}
j^v_{J,P}(x)=\epsilon_{ijk}
[q^{iT}(x)C\Gamma_{J,P}\tau q^j(x)]\Gamma^\prime_{J,P} h^k_v(x)~,
\end{equation}
where $i,~j,~k$ are the color indices, $C$ is the charge conjugation matrix, 
and $\tau$ is the isospin matrix while $q(x)$ is a light quark field.
$\Gamma_{J,P}$ and $\Gamma^\prime_{J,P}$ are some gamma matrices which 
describe the structure of the baryon with spin-parity $J^P$.  Usually $\Gamma$ 
and $\Gamma^\prime$ with least number of derivatives are used in the QCD sum 
rule method.  The sum rules then have better convergence in the high energy 
region and often have better stability.  For the ground state heavy baryon, we 
use $\Gamma_{1/2,+}=\gamma_5$, $\Gamma^\prime_{1/2,+}=1$.  In the previous 
work \cite{PLB476}, two kinds of interpolating fields are introduced to
represent the excited heavy baryon.  In this work, we find that only the 
interpolating field of transverse derivative is adequate for the analysis.
Nonderivative interpolating field results in a vanishing perturbative
contribution.  The choice of $\Gamma$ and $\Gamma^\prime$ with derivatives for 
the $\Lambda_{c1}^{1/2}$ and $\Lambda_{c1}^{3/2}$ is then 
\begin{eqnarray}
\Gamma_{1/2,-}=(a+b\fmslash v)\gamma_5~,~~~~~ 
&&\Gamma^\prime_{1/2,-}=
\frac{i{\overleftarrow {\fmslash D}_t}}{M}\gamma_5~,\nonumber\\
\Gamma_{3/2,-}=(a+bv\hspace{-1.7mm}/)\gamma_5~,~~~~~
&&\Gamma^\prime_{3/2,-}=\frac{1}{3M}(i{\overleftarrow D}^\mu_t
+i{\overleftarrow {\fmslash D}_t}\;\gamma^\mu_t)~,
\label{deriv}
\end{eqnarray}
where a transverse vector $A^\mu_t$ is defined to be 
$A^\mu_t\equiv A^\mu-v^\mu v\cdot A$, and $M$ in Eq. (\ref{deriv}) is some 
hadronic mass scale.  $a$, $b$ are arbitrary numbers between 0 and 1.

The baryonic decay constants in the HQET are defined as follows,
\begin{mathletters}
\begin{eqnarray}
\langle 0|j^v_{1/2,+}|\Lambda_{b}\rangle &=&f_{\Lambda_b}\psi_v~,\\
\langle 0|j^{v}_{1/2,-}|\Lambda_{c1}^{1/2}\rangle &=&f_{1/2}
\psi^{1/2}_{v}~,\\
\langle 0|j^{v\mu}_{3/2,-}|\Lambda_{c1}^{3/2}\rangle &=&
\frac{1}{\sqrt{3}}f_{3/2}\psi^{3/2\mu}_{v}\;,
\end{eqnarray}
\end{mathletters}
where $f_{1/2}$ and $f_{3/2}$ are equivalent since $\Lambda_{c1}^{1/2}$ and 
$\Lambda_{c1}^{3/2}$ belong to the same doublet with $j_\ell^P=1^-$. The 
QCD sum rule calculations give \cite{sumrule} 
\begin{equation}
f_{\Lambda_b}^2 e^{-{\bar\Lambda}/T}=\frac{1}{20\pi^4}
\int_0^{\omega_c}d\omega \omega^5 e^{-\omega/T}
+\frac{1}{6}\langle{\bar q}q\rangle^2 e^{-m_0^2/8T^2}
+\frac{\langle\alpha_s GG\rangle}{32\pi^3}T^2~,
\end{equation}
and \cite{PLB476}
\begin{eqnarray}
f_{1/2}^2 e^{-{\bar\Lambda}^\prime/T^\prime}&=&
\int_0^{\omega_c^\prime} d\omega \frac{3N_c!}{4\pi^4\cdot 7!}\omega^7
(24a^2+40b^2)e^{-\omega/T^\prime}+\frac{\langle\alpha_s GG\rangle}{32\pi^3}
T^{\prime 4}(-a^2+b^2)\nonumber\\
&&+\frac{N_c!}{2\pi^2}\Big[\langle{\bar q}q\rangle T^{\prime 5}(16ab)
 -\langle{\bar q}g\sigma\cdot G q\rangle T^{\prime 3}ab\Big]
-\frac{\langle{\bar q}g\sigma\cdot G q\rangle}{4\pi^2}T^{\prime 3}(3ab)~.
\end{eqnarray}
In the above equations, $T^{(\prime)}$ are the Borel parameters and
$\omega_c^{(\prime)}$ are the continuum thresholds, and $N_c=3$ is the
color number.  In the heavy quark limit, the mass parameters $\bar\Lambda$ 
and $\bar\Lambda^\prime$ are defined as
\begin{equation}
{\bar\Lambda}^\prime=M_{\Lambda_{Q1}}-m_Q~,~~~~~
{\bar\Lambda}=M_{\Lambda_Q}-m_Q~.
\end{equation}

In order to get the QCD sum rule for the IW function, one studies the 
analytic properties of the three-point correlators
\begin{mathletters}\label{3-correlator}
\begin{eqnarray}
\Xi^\mu(\omega,\omega',y)&=&i^2\int d^4x d^4z~e^{i(k^\prime\cdot x-k\cdot z)}
 \langle 0|{\cal T}~j^{v^\prime}_{1/2,-}(x)~
  {\bar h}^{(c)}_{v^\prime}(0)\Gamma^\mu h^{(b)}_v(0)
  ~{\bar j}^v_{1/2,+}(z)|0\rangle\nonumber\\&& \quad
=\Xi(\omega,\omega',y)\;(\fmslash v+y)\gamma_5\;\frac{1+\fmslash v'}
{2}\Gamma^\mu\frac{1+\fmslash v}{2}\;,\\
\Xi^{\alpha\mu}(\omega,\omega',y)&=&i^2\int d^4x d^4z~
e^{i(k^\prime\cdot x-k\cdot z)}
\langle 0|{\cal T}~j^{v^\prime\alpha}_{3/2,-}(x)~
  {\bar h}^{(c)}_{v^\prime}(0)\Gamma^\mu h^{(b)}_v(0)
  ~{\bar j}^v_{1/2,+}(z)|0\rangle\nonumber\\&& \quad
=\Xi(\omega,\omega',y)\;[(-v^\alpha+yv^{'\alpha}
+\frac{1}{3}(\gamma^\alpha+v^{'\alpha})
\;(\fmslash v-y)]\;\frac{1+\fmslash v'}{2}\Gamma^\mu\frac{1+\fmslash v}{2}\;,
\end{eqnarray}
\end{mathletters}
where $\Gamma^\mu=\gamma^\mu$ or $\gamma^\mu\gamma_5$.  The variables $k$, 
$k'$ denote residual ``off-shell" momenta which are related to the momenta 
$P$ of the heavy quark in the initial state and $P'$ in the final state by 
$k=P-m_Qv$, $k'=P'-m_{Q'}v'$, respectively.

The coefficient $\Xi(\omega,\omega',y)$ in (\ref{3-correlator}) is an 
analytic function in the ``off-shell energies" $\omega=v\cdot k$ and 
$\omega'=v'\cdot k'$ with discontinuities for positive values of these 
variables. It furthermore depends on the velocity transfer $y=v\cdot v'$, 
which is fixed at its physical region for the process under consideration. By 
saturating with physical intermediate states in HQET, one finds the hadronic 
representation of the correlators  as following
\begin{equation}
\Xi_{\rm hadron}(\omega,\omega',y)=\frac{f_{1/2}f^*_{\Lambda_b}\xi(y)}
    {\sqrt{3}({\bar\Lambda}^\prime-\omega^\prime)({\bar\Lambda}-\omega)}
    +\mbox{higher resonances} \;.
\label{hadronic}
\end{equation}
In obtaining above expression  the Dirac and Rartia-Schwinger spinor sums
\begin{eqnarray}
&& \Lambda_+=\sum_{s=1}^2u(v,s)\bar u(v,s)={1+\fmslash v\over 2}\nonumber\\
&& \Lambda_{+}^{\mu\nu}=\sum_{s=1}^4u^\mu(v,s)\bar u^\nu(v,s)=(-g_t^{\mu\nu}+
\frac{1}{3}\gamma_t^\mu\gamma_t^\nu){1+\fmslash v\over 2}\;
\end{eqnarray}
have been used, where $g^{\mu\nu}_t=g^{\mu\nu}-v^\mu v^\nu$. 

In the quark-gluon language, $\Xi(\omega,\omega',y)$ in 
Eq. (\ref{3-correlator}) is written as
\begin{eqnarray}
\Xi(\omega,\omega',y)=\int^\infty_0 d\nu d\nu^\prime 
\frac{\rho^{\rm pert}(\nu,\nu^\prime,y)}
 {(\nu-\omega)(\nu^\prime-\omega^\prime)}+({\rm subtraction})
 +\Xi^{\rm cond}(\omega,\omega',y)\;,
\label{quarkgluon}
\end{eqnarray}
where the perturbative spectral density function 
$\rho^{\rm pert}(\nu,\nu^\prime,y)$ and the condensate contribution 
$\Xi^{\rm cond}$ are related to the  calculation of the Feynman diagrams 
depicted in Fig. 1.

The QCD sum rule is obtained by equating the phenomenological and 
theoretical expressions for $\Xi$. In doing this the quark-hadron duality 
needs to be assumed to model the contributions of higher resonance part of 
Eq. (\ref{hadronic}). Generally speaking, the duality is to simulate the 
resonance contribution by the perturbative part above some thresholds 
$\omega_c$ and $\omega'_c$, that is
\begin{equation}
{\rm res.}=\int^\infty_{\omega_c}\int^\infty_{\omega^\prime_c}d\nu 
 d\nu^\prime~\frac{\rho^{\rm pert}(\nu,\nu^\prime,y)}
 {(\nu-\omega)(\nu^\prime-\omega^\prime)}~.
\label{res}
\end{equation} 
In the QCD sum rule analysis for  $B$ semileptonic decays into ground state 
$D$ mesons, it was argued by Neubert in \cite{Neubert}, and Blok and Shifman 
in \cite{Shifman} that the perturbative and the hadronic spectral densities 
can not be locally dual to each other, the necessary way to restore duality 
is to integrate the spectral densities over the ``off-diagonal'' variable 
$\nu_-=\sqrt{\frac{y+1}{y-1}}(\nu-\nu')/2$, keeping the ``diagonal'' variable 
$\nu_+=(\nu+\nu')/2$ fixed. It is in $\nu_+$ that the quark-hadron duality is 
assumed for the integrated spectral densities.  The same prescription shall be 
adopted in the following analysis.  On the other hand, in order to suppress the 
contributions of higher resonance states a double Borel transformation in 
$\omega$ and $\omega'$ is performed to both sides of the sum rule, which 
introduces two Borel parameters $T_1$ and $T_2$.  For simplicity we shall take 
the two Borel parameters equal: $T_1 = T_2 =2T$. 

Combining Eqs. (\ref{hadronic}), (\ref{quarkgluon}), our duality assumption 
and making the double Borel transformation, one obtains the sum rule for 
$\xi(y)$ as follows
\begin{equation}
{f_{1/2}f^*_{\Lambda_b}\xi(y)\over\sqrt 3}
e^{-({\bar\Lambda}^\prime+{\bar\Lambda})/2T}=
2\;\Bigg(\frac{y-1}{y+1}\Bigg)^{1/2}\int^{\omega_c(y)}_0 d\nu_+e^{-\nu_+/T}
\int^{\nu_+}_{-\nu_+}d\nu_-\rho(\nu_+,\nu_-;y)
+{\hat B}^{\omega^\prime}_{2T}{\hat B}^\omega_{2T}\Xi^{\rm cond}~,
\label{sumrule}
\end{equation}
where $\nu=\nu_++\sqrt{\frac{y-1}{y+1}}\nu_-$, 
$\nu^\prime=\nu_+-\sqrt{\frac{y-1}{y+1}}\nu_-$.

Confining us to the leading order of perturbation and the operators with 
dimension $D\leq 6$ in OPE, the spectral density 
$\rho^{\rm pert}(\nu,\nu^\prime;y)$ and 
${\hat B}^{\omega^\prime}_{2T}{\hat B}^\omega_{2T}\Xi^{\rm cond}$ are 
\begin{eqnarray}
\rho(\nu,\nu^\prime;y)&=&
 \frac{36a}{\pi^4}\frac{1}{2!3!}\Bigg(\frac{1}{2\sqrt{y^2-1}}\Bigg)^7
\nonumber\\
&&\times \Big[A(\nu,\nu^\prime;y)^3B(\nu,\nu^\prime;y)^2
   -A(\nu,\nu^\prime;y)^2B(\nu,\nu^\prime;y)^3\Big]\nonumber\\
{\hat B}^{\omega^\prime}_{2T}{\hat B}^\omega_{2T}\Xi_{(b)}&=&0~,\nonumber\\
{\hat B}^{\omega^\prime}_{2T}{\hat B}^\omega_{2T}\Xi_{(d)}&=&
  -\frac{b}{48\pi^2}\langle{\bar q}g\sigma\cdot Gq\rangle(2T)^2
     \frac{2y+1}{(1+y)^2}~,\nonumber\\
{\hat B}^{\omega^\prime}_{2T}{\hat B}^\omega_{2T}
 \big\{\Xi_{(c)}+\Xi_{(e)}\big\}
  &=&\frac{b}{2\pi^2}\frac{1}{(1+y)^2}\Bigg[
   2\langle{\bar q}q\rangle(2T)^4
   -\langle{\bar q}g\sigma\cdot G q\rangle(2T)^2\frac{4y+5}{48}\Bigg]~,
\nonumber\\
{\hat B}^{\omega^\prime}_{2T}{\hat B}^\omega_{2T}
 \big\{\Xi_{(f)}+\Xi_{(g)}+\Xi_{(h)}\big\}&=&
  -\frac{a}{192\pi^3}\langle\alpha_s GG\rangle T^3\frac{-20y+67}{(1+y)^3}~,
\label{result}
\end{eqnarray}
where 
\begin{eqnarray}
A(\nu,\nu^\prime;y)&=&
  \Bigg(\nu_+ -\sqrt{\frac{y-1}{y+1}}\nu_-\Bigg)e^\theta
 -\Bigg(\nu_+ +\sqrt{\frac{y-1}{y+1}}\nu_-\Bigg)~,\nonumber\\
B(\nu,\nu^\prime;y)&=&
 \Bigg(\nu_+ +\sqrt{\frac{y-1}{y+1}}\nu_-\Bigg)
 -\Bigg(\nu_+ -\sqrt{\frac{y-1}{y+1}}\nu_-\Bigg)e^{-\theta}~,\nonumber\\
\sinh\theta&=&\sqrt{y^2-1}~.
\end{eqnarray}
Here the dimensionful parameter $M$ in Eq. (\ref{deriv}) is dropped since 
it cancels out in (\ref{sumrule}).


\section{Results and discussion}

For the numerical analysis, the standard values of the condensates are used;
\begin{eqnarray}
\langle{\bar q}q\rangle&=&-(0.23~{\rm GeV})^3~,\nonumber\\
\langle\alpha GG\rangle&=&0.04~{\rm GeV}^4~,\nonumber\\
\langle \bar{q}g\sigma\cdot G q\rangle&\equiv&m_0^2\langle{\bar q}q
\rangle~,~~~~~m_0^2=0.8~{\rm GeV}^2~.
\end{eqnarray}
In dealing with the variables, some remarks should be noticed.  First, the 
continuum threshold $\omega_c^\prime$ in $f_{\frac{1}{2}(\frac{3}{2})}$
(${\bar\Lambda}^\prime$) can differ from that in $f_{\Lambda_b}$ 
(${\bar\Lambda}$).  However, it is expected that the values of $\omega_c$ and 
$\omega_c'$ have no significant difference.  This is because the mass 
difference ${\bar\Lambda}^\prime-{\bar\Lambda}$ is not large \cite{PLB476},
${\bar\Lambda}^\prime-{\bar\Lambda}\simeq 0.2~{\rm GeV}$.  Indeed the central 
values of them were close to each other in the sum rules analysis for 
$f_{\frac{1}{2}(\frac{3}{2})}$ (${\bar\Lambda}^\prime$) and $f_{\Lambda_b}$ 
(${\bar\Lambda}$).  In addtion, the continuum threshold $\omega_c(y)$ in 
Eq. (\ref{sumrule}) in general can be a function of $y$.  We take it to be a 
constant $\omega_c(y)=\omega_c=\omega_c'=\omega_0$ in the numerical analysis.  
In this sense, we use only one continuum threshold throughout the analysis.
Second, there are input parameters of $a$ and $b$ in the interpolating fields
(\ref{deriv}).  In \cite{PLB476}, the choice of $(a,b)=(1,0)$ shows the best 
stability for the mass parameter ${\bar\Lambda}^\prime$.  We adopt the same 
set of $(a,b)=(1,0)$ in this analysis.  Third, there are two Borel parameters 
$T_1$ and $T_2$ in general, corresponding to $\omega$ and $\omega^\prime$ in 
$\Xi(\omega, \omega',y)$, respectively.  We have taken $T_1=T_2$ in the 
analysis.  In \cite{Huang} for $B$ into excited charmed meson transition,
the authors got a $10\%$ increase of the leading IW function at zero recoil 
when $T_2/T_1=1.5$ compared to the value when $T_1=T_2$.  It seems quite 
reasonable to expect that in the heavy baryon case, the numerical results are 
similar for small variations around $T_2/T_1=1$.

The leading IW function $\xi(y)$ is plotted in Figs.\ \ref{3dplot},\ref{IW}.
In Fig.\ \ref{3dplot}, we give a three-dimensional plot of $\xi=\xi(y,T)$.
The best stability is shown within the sum rule window,
\begin{equation}
0.16\le T\le 0.6~({\rm GeV})~.
\end{equation}
The upper and lower bounds are fixed such that the condensate contribution
amounts to at most $30\%$ while the pole contribution to $50\%$.  Note that 
this range has overlaps with the sum rule windows in \cite{PLB476} and 
\cite{sumrule}.  This reflects the self-consistence of the sum rule analysis.  
In Fig.\ \ref{IW}, the band corresponds to the variation of $\xi(y)$ from
$\omega_0=1.2$ to $\omega_0=1.6$ GeV.  In addition, we have found that there 
is almost no numerical difference if the threshold $\omega_c(y)$ is instead 
taken to be $(1+y)\omega_0/2y$ which was suggested in \cite{Neubert}.  This 
is because the allowed kinematical region is very narrow around $y\simeq 1$.
At zero recoil, $\xi(1)$ is
\begin{equation}
\xi(1)=0.29_{-0.035}^{+0.038}~,~~~{\rm for}~~~\omega_0=1.4\pm 0.1~{\rm GeV}~,
\end{equation}
and the slope parameter $\rho^2$ in (\ref{rho}) for different $\omega_0$ is 
\begin{equation}
\rho^2=2.01^{+0.003}_{-0.005}~,~~~{\rm for}~~~\omega_0=1.4\pm 0.1~{\rm GeV}~.
\end{equation}
This value is somewhat larger than the large $N_c$ HQET prediction in 
\cite{Leibovich}.

\section{Summary}

For the weak decays of the $\Lambda_b$ baryon to the excited charmed 
baryons $\Lambda_{c1}^{1/2,3/2}$, by using QCD sum rules, we have obtained 
the information of the leading IW function which has been defined in Eqs. 
(\ref{formfactor}) and (\ref{formfactors}), within the framework of HQET,
\begin{equation}
\xi(y)=0.29[1-2.01(y-1)]~.
\end{equation}
The sum rule uncertainty of $\xi(y)$ can be found in Eqs. (23) and (24).  
Compared to the result of the large $N_c$ HQET \cite{Leibovich}, the main 
difference here lies in the value of $\xi(1)$.  The branching ratios are 
therefore estimated to be smaller than those given in \cite{Leibovich},
\begin{equation}
{\rm Br}.(\Lambda_b\to\Lambda_{c1}^{1/2,3/2}e\bar{\nu}_e)
\simeq 0.21-0.28\%~.
\end{equation}
The future experiments will check this prediction.

\begin{center}
{\large\bf Acknowledgment}
\end{center}

We would like to thank I. W. Stewart for helpful communication.  
This work was supported in part by the BK21 program of Korea, and the 
National Natural Science Foundation of China.  


\newpage

\begin{center}{\large\bf FIGURE CAPTIONS}\end{center}

\noindent
Fig.~1
\\
Feynman diagrams for the three-point function with derivative 
interpolating fields.  Double line denotes the heavy quark.
\vskip .3cm
\par
\noindent
Fig.~2
\\
Three-dimensional plot of IW function for $1\le y\le 1.2$ and
$0\le T\le 1$ (GeV).  The continuum threshold is chosen to be 
$\omega_c(y)=1.4$ GeV.
\vskip .3cm
\par
\noindent
Fig.~3
\\
IW function as a function of $y$ for various $\omega_0$ at fixed 
$T=0.38$ GeV.
The lowest line corresponds to $\omega_0=1.2$ GeV while the highest to 
$\omega_0=1.6$ GeV, with the increment $0.1$ GeV.


\begin{figure}
\vskip 2cm
\begin{center}
\epsfig{file=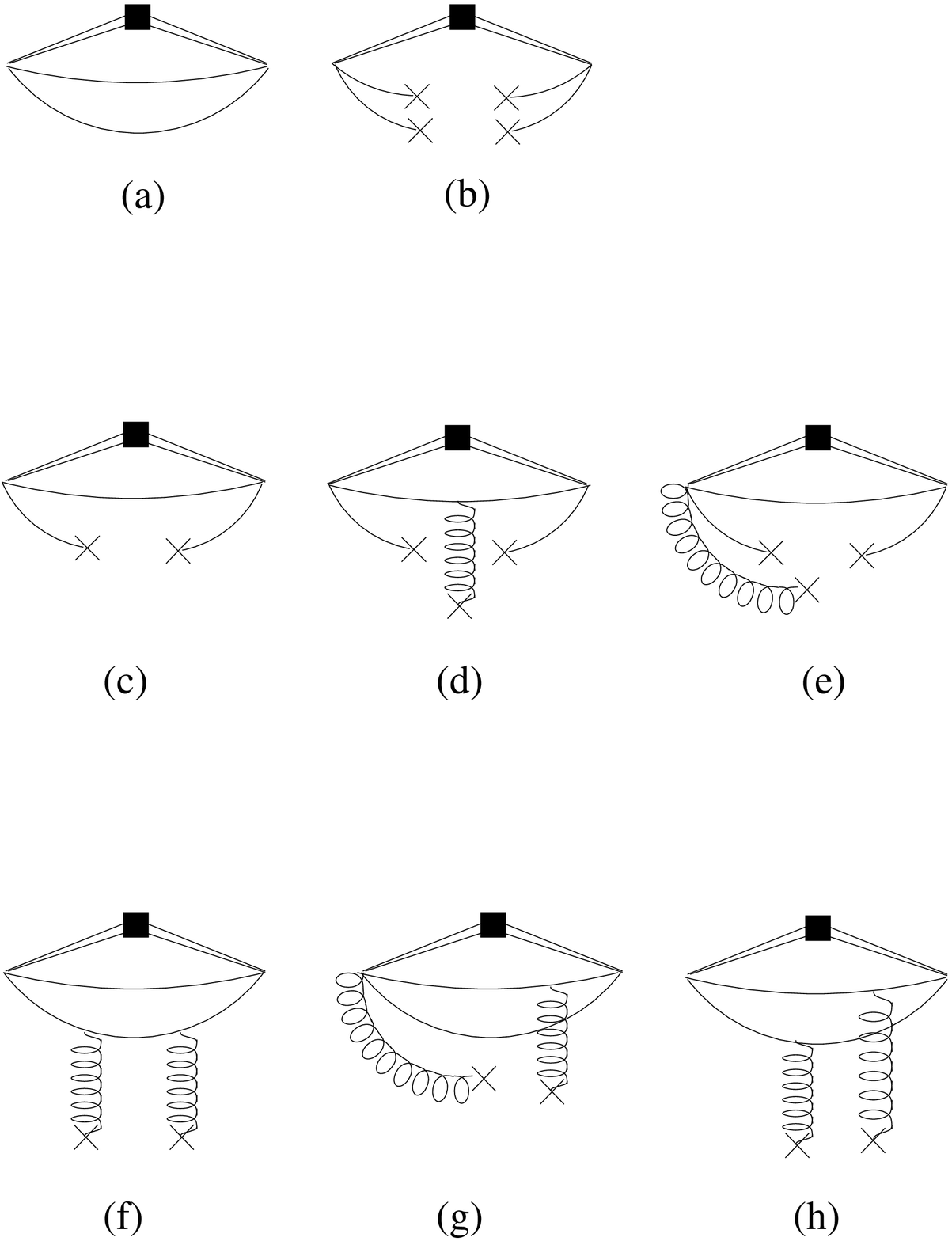, height=15cm}
\end{center}
\caption{}
\label{diagram}
\end{figure}

\pagebreak

\begin{figure}
\begin{center}
\epsfig{file=3d.epsi, height=7cm}
\end{center}
\caption{}
\label{3dplot}
\end{figure}
\pagebreak
\begin{figure}
\begin{center}
\epsfig{file=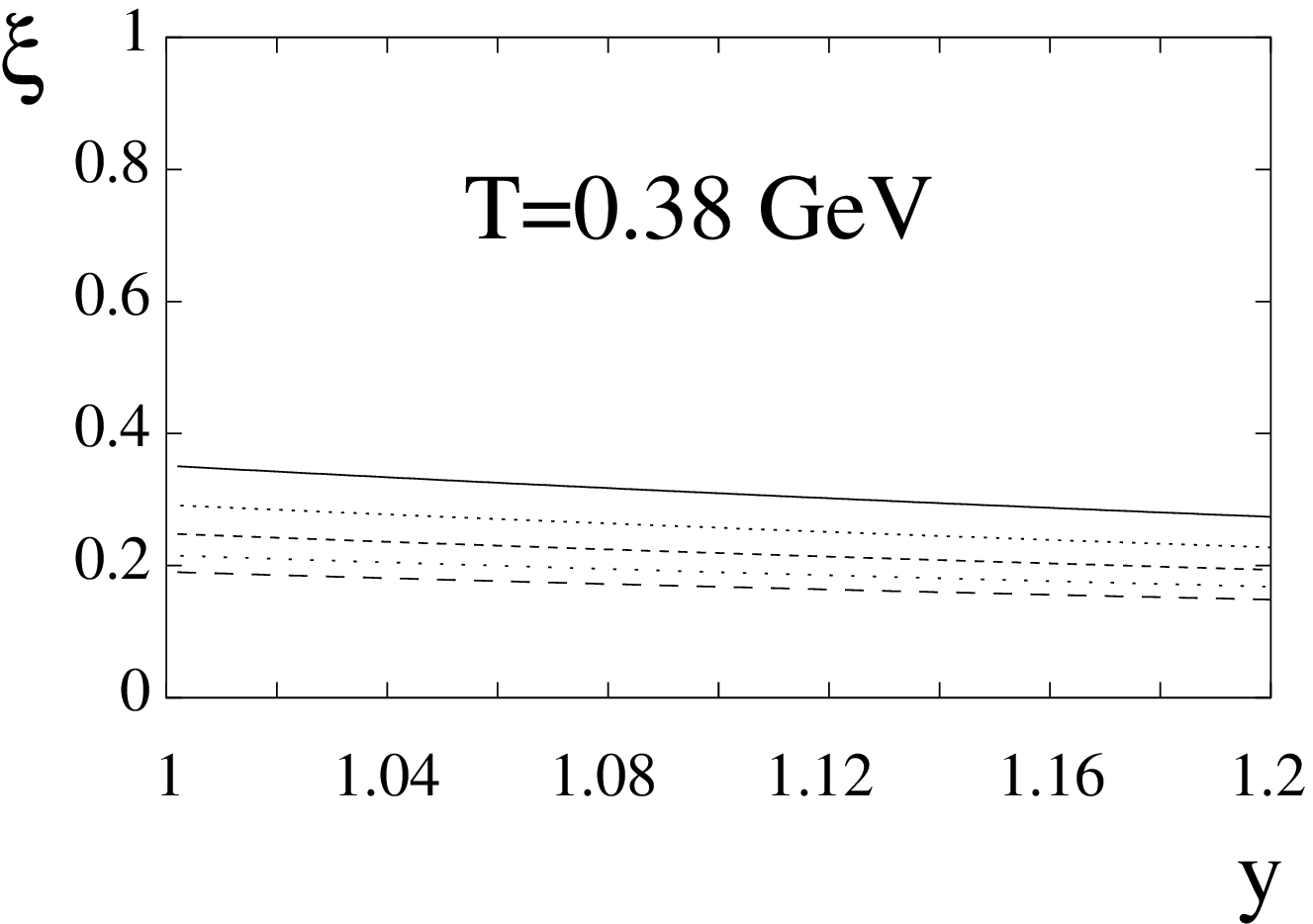, height=7cm}
\end{center}
\caption{}
\label{IW}
\end{figure}

\end{document}